\documentclass[%
 reprint,
 amsmath,amssymb,
 aps,
 prl,
]{revtex4-1}

\usepackage[version=3]{mhchem} 
\usepackage[T1]{fontenc}
\usepackage{textgreek}  
\usepackage{color}
\usepackage{graphicx}
\usepackage{dcolumn}
\usepackage{bm}
\usepackage[table,xcdraw]{xcolor}

\usepackage{amsmath}
\usepackage{footnote}
\usepackage{mathtools}

\newcommand{\RNum}[1]{\uppercase\expandafter{\romannumeral #1\relax}}

\usepackage{array,multirow}

\begin{document}

\title{Data Enhanced Reaction Predictions in Chemical Space With Hammett's Equation
}

\author{Marco Bragato}
\author{Guido Falk von Rudorff}
\author{O. Anatole von Lilienfeld}
\email{anatole.vonlilienfeld@unibas.ch}
\affiliation{Institute of Physical Chemistry and National Center for Computational Design and Discovery of Novel Materials (MARVEL) Department of Chemistry, University of Basel, Klingelbergstrasse 80, CH-4056 Basel, Switzerland}

\begin{abstract}    
By separating the effect of substituents from chemical process variables, such as reaction mechanism, solvent, or temperature, the Hammett equation enables control of chemical reactivity throughout chemical space.
We used global regression to optimize Hammett parameters $\rho$ and $\sigma$
in two datasets, experimental rate constants for benzylbromides reacting with thiols and the decomposition of ammonium salts, and a synthetic dataset consisting of computational activation energies of {\raise.17ex\hbox{$\scriptstyle\mathtt{\sim}$}}1400 $S_N2$ reactions, with various nucleophiles and leaving groups (-H, -F, -Cl, -Br) and functional groups (-H, -NO$_2$, -CN, -NH$_3$, -CH$_3$).
The original approach is generalized to predict potential energies of activation in non aromatic molecular scaffolds with multiple substituents.
Individual substituents contribute additively to molecular $\sigma$ with a unique regression term, which quantifies the inductive effect.
Moreover, the position dependence of the substituent can be replaced by a distance decaying factor for $S_N2$. 
Use of the Hammett equation as a base-line model for $\Delta$-Machine learning models of the activation energy in chemical space results in substantially improved learning curves for small training set sizes. 
\end{abstract}

\maketitle

\section{Introduction}\label{intro}
Chemical reactions are difficult to study and model from a theoretical point of view. In 1935, Hammett proposed a quantitative model for free energy differences in benzyl derivatives\cite{Hammett1937,Hammett1935} that 
assumes that the substituent and reaction effects can be separated by a product {\em Ansatz}:
\begin{equation}
	\log \left ( \frac{K}{K_0} \right ) \simeq \rho \sigma
	\label{VanillaHamm}
\end{equation}
Here, $K$ is either the equilibrium or rate constant for a substituted reactant, $K_0$ refers to the unsubstituted reactant, $\rho$ is a constant that depends only on the reaction, taking into account also conditions such as temperature and solvent and $\sigma$ depends only on the type of substituent and its position on the molecule.

This model is compelling since it gives an intuitive concept of electron donating and electron withdrawing effects\cite{Jaffe1953,Krygowski2005,Exner1999,Cherkasov1998} in the context of free energy differences. 
The model quickly became quite successful and has been applied to problems ranging from its original purpose, quantifying substituent effects\cite{Jaffe1953}, to redox potentials\cite{Masui1993}, dipole moments \cite{Beek1957}, orbital energies of metallorganic complexes \cite{Chang2019}, aromaticity \cite{Szatylowicz2017,Stasyuk2016,Szatylowicz2017-2,Stasyuk2016,Szatylowicz2017-2,Szatylowicz2017-3,Gershoni-Poranne2018,Hansch1973,Katritzky1977,DiLabio1999, DiLabio2000,Palat2001,Krygowski2004,Dey2019}, 
ion stabilization \cite{Buszta2019}, mechanicistic investigation \cite{Cruz2019,Barbee2018}, catalyst activity of nanoparticles \cite{Kumar2018}, proton-electron coupling in radicals \cite{Kimura2019}, molecular conductance \cite{Venkataraman2007}, excited singlet state \cite{Dobrowolski2018}, and even toxicities\cite{Song2006}.
More recent approaches have also tried to apply the models to non-benzyl systems \cite{Liveris1956, Chang2019, Ayoubi-Chianeh2019, Kilde2017}.
It is, however, less satisfying because the linear relationship postulated by Hammett lacks a motivation based on physical effects. Early attempts to explain the theory by electrostatic considerations\cite{Gallup1952,Price1941} were successful for special cases only. 
Nevertheless, Hammett's model has demonstrated remarkable predictive power and accuracy for many cases given the model's simplicity\cite{Jaffe1953}. 
Over time the equation has been expanded to also encompass, solvent effect \cite{Jahagirdar1988, Kondo1969, Grunwald1948, Winstein1951}, resonance and field effect \cite{Swain1968},  steric effects \cite{Taft1952, Taft1952-2, Taft1953,Santiago2016}, nucleophilicity \cite{Swain1953} and oxidation potential \cite{Edwards1954}.
These models trade off transferability for accuracy; for this reason, in the majority of applications, the original equation is the one being used.

With the promise of Hammett's model that substituent effects can be separated from other contributions to a reaction rate, a certain transferability of the substituent parameters $\sigma$ seems to be guaranteed.
However, it is hard to assign unambiguous values of $\sigma$ to functional groups, as they often lack transferability, such that the reference reaction and compound becomes of utmost importance\cite{Pearson1952}.
Similarly, $\rho$ has shown to be hardly transferable and even exhibit an inconsistent temperature dependence\cite{Jaffe1953}.

Interestingly, Hammett parameters can be inferred from experiments: either by OH vibrational frequencies related to the electron density at the point of bonding\cite{Baker1959}, by assessing NMR shifts\cite{Yoder1969,Axenrod1969,Taft1960,Thirunarayanan2007} or quadrupole resonance\cite{Bray1957,Bray1954}, by relation to electron binding energies\cite{Lindberg1976,Takahata2005}, IR spectroscopy \cite{Liler1965}, electrochemical polarization \cite{Sarkar2019}, or charge transfer\cite{Star2003}.
Extensive comparison to experiment however, uncovered special cases in which Hammett's model struggles to adequately model reality, partially leading to the introduction of several $\sigma$ values for the same functional group to be used in different molecular environments\cite{Huenig1953}. 
Some limitations subsequently could be surpassed by extending the model, e.g. to include concentration dependence\cite{Hansch1962}. 

From a computational perspective, atomic charges were quickly found to correlate with $\sigma$ values for a given functional group\cite{Ertl1997,Larsen1975,Genix1996}, so the few available experimental data points that otherwise would be tedious to extend could be used to calibrate a linear regression while the functional groups were quickly screened by simple charge fitting methods or electron density self-similarity measures\cite{Girones2006}. 
Still, the resulting $\sigma$ values lack transferability\cite{Hine1959} and computational studies were not successful for reactions involving excited states\cite{Wagner1976}. 
More recently, energy decomposition approaches have been evaluated\cite{Fernandez2006}, connecting to the idea of electrostatic contributions as a dominating contribution to the validity of Hammett's model.

The use of Hammett's approach as a guide in chemical space to find molecules of desired energy differences has been hampered by three issues: the focus on single substituents, the difficulty to obtain a consistent set of Hammett coefficients\cite{Jaffe1953,Lichtin1952,White1961} and the restriction to free energy differences.
While multiple substituents have been cautiously explored\cite{Shorter1949}, experimental evidence was found that $\sigma$ values of multiple substituents are additive, as long as no resonance is involved\cite{Yukawa1959,Taft1957,Cherkasov1998}. 
In this work, we focus on addressing these three main limitations of Hammett's approach.

\section{Method}
\subsubsection{The Hammett equation}\label{method:Hamm}
The original formulation of the Hammett equation is shown at the beginning of section \ref{intro}. 
Here the only observables are the reaction constants $K$ and $K_0$, so it is not possible to calculate a unique set of $\{\rho\}$ and $\{\sigma\}$, as there will always be an arbitrary constant that can be moved between the two. 
In order to remove this degree of freedom, Hammett proposed the following procedure~\cite{Hammett1937}: (i) pick a reference reaction $i$ for which $\rho_i \coloneqq 1$, (ii) use it to assign a value of $\sigma$ to the substituents for which there is data for the reference reaction, (iii) use this set $\{\sigma\}$ to evaluate $\rho_j$ for another reaction $j$ using a least squares regression, (iv) expand the set $\{\sigma\}$ using the new $\rho_j$, (v) repeat steps (iii) and (iv) until each reaction and substituent has a value assigned.

The choice of the reference reaction, as well as the sequence used to expand the set $\{\sigma\}$, greatly influences the final result: for a set of $N_R$ reactions there are up to $N_R!$ possible sets of $\{\rho\}$ and $\{\sigma\}$.
Overall, with $N_R$ reactions and $N_S$ set of substituents there are $N_R  N_S$ different Hammett equations with only $N_R+N_S$ parameters to determine. 
The system is greatly overdetermined, making it is easy to overfit of the model.

In our model, we use a robust regressor~\cite{Theil1950} to limit the influence of outliers, and we calculate the entire set of reaction constants $\{\rho\}$ at once, thus removing the dependence on the choice of the reference reaction. 
The substituent constants $\{\sigma\}$ are then evaluated by inverting the Hammett equation and averaging the results over all reactions.
For numerical reasons, it might be necessary to initially fix one arbitrary reaction constant to 1 to avoid trivial solutions.
This is the only source of bias in the model, meaning that the number of possible set of reaction and substituent constant scales only linearly with the number of reactions, and not factorially like in the original model.
This procedure allows to affordably identify the best set of parameters.
The derivation of the model is explained in details in the Supplementary Information.

For reactants with multiple substituents, $\sigma$ describes the combined effect of all of them.
To identify individual contributions, we propose a linear model where the molecular $\sigma$ is given by the sum of single substituent parameters $\tilde{\sigma}$, obtained by a categorical regression using a dummy encoding.
These term depend on the chemical composition of the substituent and on its position on the molecule.
In order to separate these two contributions, we modelled each single substituent constant as a product between a term $\alpha$, which depends only the chemical composition, and a distance decaying function (exponential or power law), which encodes the distance of the substituent from the reaction center. 

To distinguish the two methods of calculating the substituent constants, i.e. by reversing the Hammett equation and by summing single substituents contributions,we named the first one $\sigma$-Hammett and the latter $\alpha$-Hammett.

Non-linear functions, which can model many body contributions, have also been studied by including three body terms such as the Axilrod-Teller-Muto\cite{ATM} potential. 
This increases the number of parameters needed but allows to include the interactions between substituents.

\subsubsection{Machine learning}
We trained a Kernel Ridge Regression (KRR) machine to learn the kinetic constant and activation energies for different reactions.
Molecules were described with a one-hot encoding representation, which maps every fragment into a fingerprint-like string of zeroes and ones.
Our Hammett model was then used as a baseline for Delta Machine Learning ($\Delta$-ML), where a machine was trained to learn the residuals of the method.
This approach can give a faster learning, since the hypersurface of the residuals is usually smoother, thus easier to learn.

These models were programmed in Python using the QML~\cite{QMLcode} and scikit-learn~\cite{scikit-learn} packages. 
Hyper-parameters were determined with a 5-fold validated grid search, final results obtained with a 15-fold cross validation.

\section{Results}
\subsubsection{Experimental analysis}
\begin{figure}[h!]
	    \centering
    \includegraphics[scale=0.4]{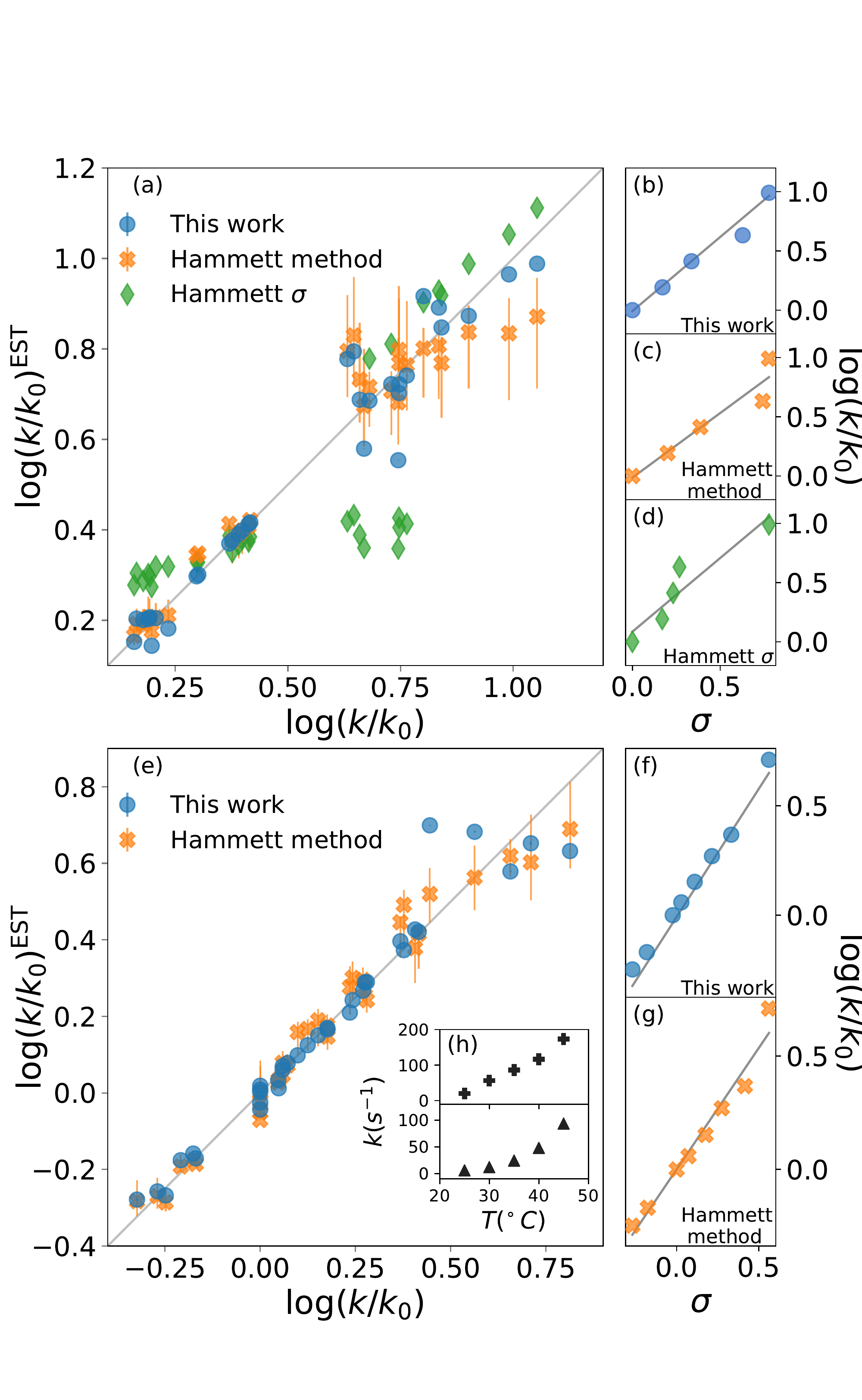}
    \caption{Prediction of kinetic constants on two experimental reaction data set: nucleophilic substitution between benzilbromides and thiols~\cite{Hudson1962}(top half) and decomposition of ammonium salts (bottom half)\cite{Burns1969}. The picture compares results from our model (blue circles), from the original Hammett procedure (orange crosses) and from the tabulated parameters of the original paper (green diamonds)~\cite{Hammett1937}.
    The correlation plots (a) and (e) show the higher reliability of our method for the prediction of the rate constants when compared to the others. The error bars display the dependence on the reference reaction. The Hammett plots on the right ((b), (c), (d), (f) and (g)) show the increased robustness of our method with respect to outliers and the preservation of the relative ordering of the subsituent constans $\sigma$. The inset (h) reports the temperature dependence of the rate constants for the decomposition of two different ammonium salts, highlighting how the outliers correspond to unphysical behaviour}
    \label{fig:experimental-data}
\end{figure}

To test the effectiveness of our method, we apply it to two different set of experimental results and compare our predictions with the one from the original Hammett model~\cite{Hammett1937}. 
The first data set~\cite{Hudson1962} studies the substituent effect on the nucloephilic reactivity between tiophenols and benzilbromides. 
The second data set~\cite{Burns1969} reports the rate constants of the decomposition of tetra-alkylammonium salts in solution at different temperatures. 
According to the original formulation of Hammett, the temperature dependence is included in eq.~\ref{VanillaHamm} through the reaction constant, meaning that each temperature is described by a different $\rho$.

The kinetic constants have been evaluated through the Hammett equation using three different set of parameters $\{\rho\}$ and $\{\sigma\}$: the first one obtained with our model, the second one by applying the original Hammett method, as described in the beginning of sect.~\ref{method:Hamm}, and the third one using the values of $\sigma$ calculated by Hammett himself in the original paper \cite{Hammett1937}. 
This last method could be used only for the first of the two experimental data set, since the molecules used in the second one where not included in the original paper.

The results are shown in figure~\ref{fig:experimental-data}.
The upper half (subplots (a) to (d)) shows the results on nuclephilic substitution of benzylbromides~\cite{Hudson1962}, while the bottom half ((e) to (i)) the ones on the ammonium salts decomposition\cite{Burns1969}. 

The scatter plots (a) and (e) present the correlation between the experimental kinetic constants and the estimated ones.
The blue dots are obtained by our model, the orange cross by the original approach~\cite{Hammett1937} and the green diamond are calculated using the $\{\sigma\}$ from the original paper~\cite{Hammett1937}. 
The error bars show the range of results spanned by changing the reference reaction.
For nuclephilic substitution of thiols (upper half), the reference uses the un-subtituted thiol, while for the thermal decomposition of ammonium salts (bottom half), the reference is the reaction at 35$^{\circ}$\,C.

The correlation plots show how our method outperforms the original Hammett method in the vast majority of the case, often by a significant margin; using the original $\{\sigma\}$ yields very inaccurate results.
The error bars demonstrate how important the choice of the reference reaction is: for our method the effect is too small to be visible, while for the original method it can give results that vary by up to 25$\%$ for both the first (a) and the second (b) data set. 
The usage of tabulated sigma removes this dependence but introduces a significant error that can be up to 50$\%$. 

The improvement given by our method is in part due to the increased robustness towards outliers. 
This effect becomes evident from the Hammett plots on the right panels ((b) to (d) and (f) and (e)), which show the linear relationship between substituent constant $\sigma$ and log(k/k$_0$) for each approach.
Our method (panels (b) and (f)) gives a better interpolation for the majority of the data.
Additionally, the Hammett plots show how the ordering of the different $\sigma$ for different substituents does not depend on the method, meaning that it is still possible to use them as a relative measure of the inductive effect without loss of generality.
This comes at the cost of a worse evaluation of the cases that deviate from the linearity.

The tradeoff in accuracy on the outliers is especially evident from the scatter plot (e) for the decomposition of ammonium salts. 
The original model gives better predictions only for some specific cases, for example when considering the reaction involving a beta-naphtyl thiol. 
The dependence of the kinetic constant of this last case on the temperature is shown in the top panel of inset (h). 
The linear behaviour is in contrast with the typical exponential Arrhenius-like that can be observed for any other case in this data set, as presented in the bottom panel of (h) for a para-Methoxy thiol.
This shows that the robustness of the revised Hammett proves useful when dealing with noisy data and can be helpful in identifying unphysical features in the data set.

\subsubsection{Hammett revisited for S$_N$2}
\begin{figure}[h!]
	    \centering
    \includegraphics[scale=0.23]{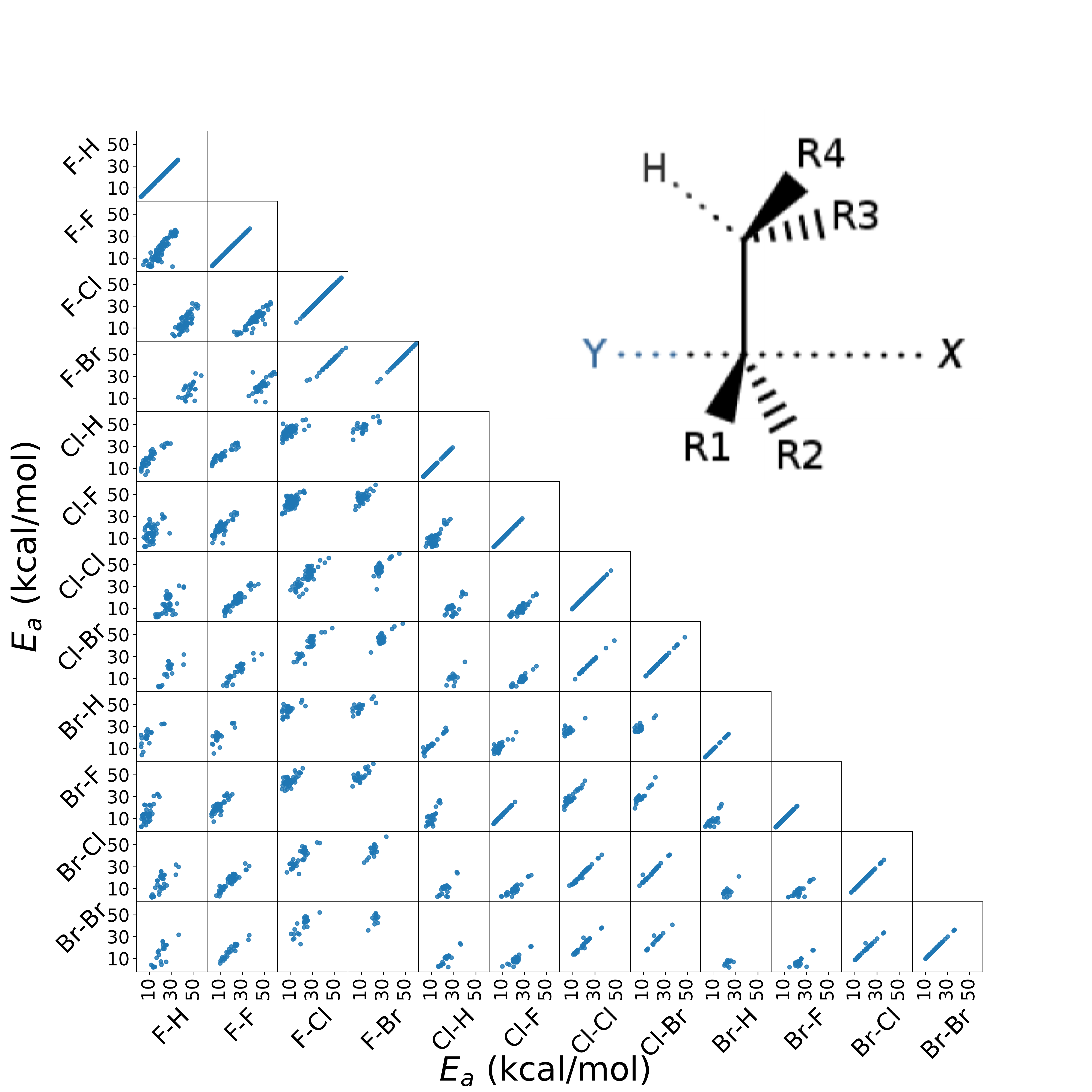}
    \caption{Correlation of the activation energies between the reactions in the data set. The labels indicate the nucleophile-leaving group couple, in this order. The data show a linear trend, which is the underlying assumption for the Hammett model. These activation energies range linearly between 3\,kcal/mol and 40\,kcal/mol. The inset in the top right corner shows the general scaffold of the molecules in the data set, where R1 to R4 are the substituents and X and Y are the nucleophile and leaving group respectively.}
    \label{fig:compareEa}
\end{figure}

In this work, we extended the Hammett equation to a chemical space that is outside the scope of the original model by working on a computational data set of S$_N$2 reactions on small molecules with an ethylene scaffold. The typical transition state is depicted in the top right inset of figure~\ref{fig:compareEa}.
These molecules have four sites where substituents can be placed, labelled R$_1$ to R$_4$, and undergo a nucleophilic substitution of the leaving group X by the nucleophile Y.
The substituents considered for positions R1 to R4 are -H, -NO$_2$, -CN, -NH$_3$, -CH$_3$, while the leaving groups and nucleophiles are: -H, -F, -Cl, -Br.
The data set was calculated by Von Rudorff et al., soon to be published.

For this data, we worked with activation energies instead of the kinetic constant.
The two quantities are related by the transition state theory, which assumes a quasi-chemical equilibrium between reactants and transition state. 
Thus, the Hammett equation can be applied to potential energy differences without loss of generality.

Activation energies for the different reactions correlate linearly with each other, as shown in the lower left part of figure~\ref{fig:compareEa}.  
Here each scatter plot compares the energy barriers of any two reactions; the nucleophile and leaving group are indicated on the edges, in this order.
The correlations ensures that the relative effect of different substituents is the same even across different reactions, which means that the ordering of the elements in $\{\sigma\}$ is univoque. 
The slope of each linear fit expresses the relative susceptibility of the two reactions to the substituents' effect. 

\begin{figure}[h!]
	    \centering
    \includegraphics[scale=0.55]{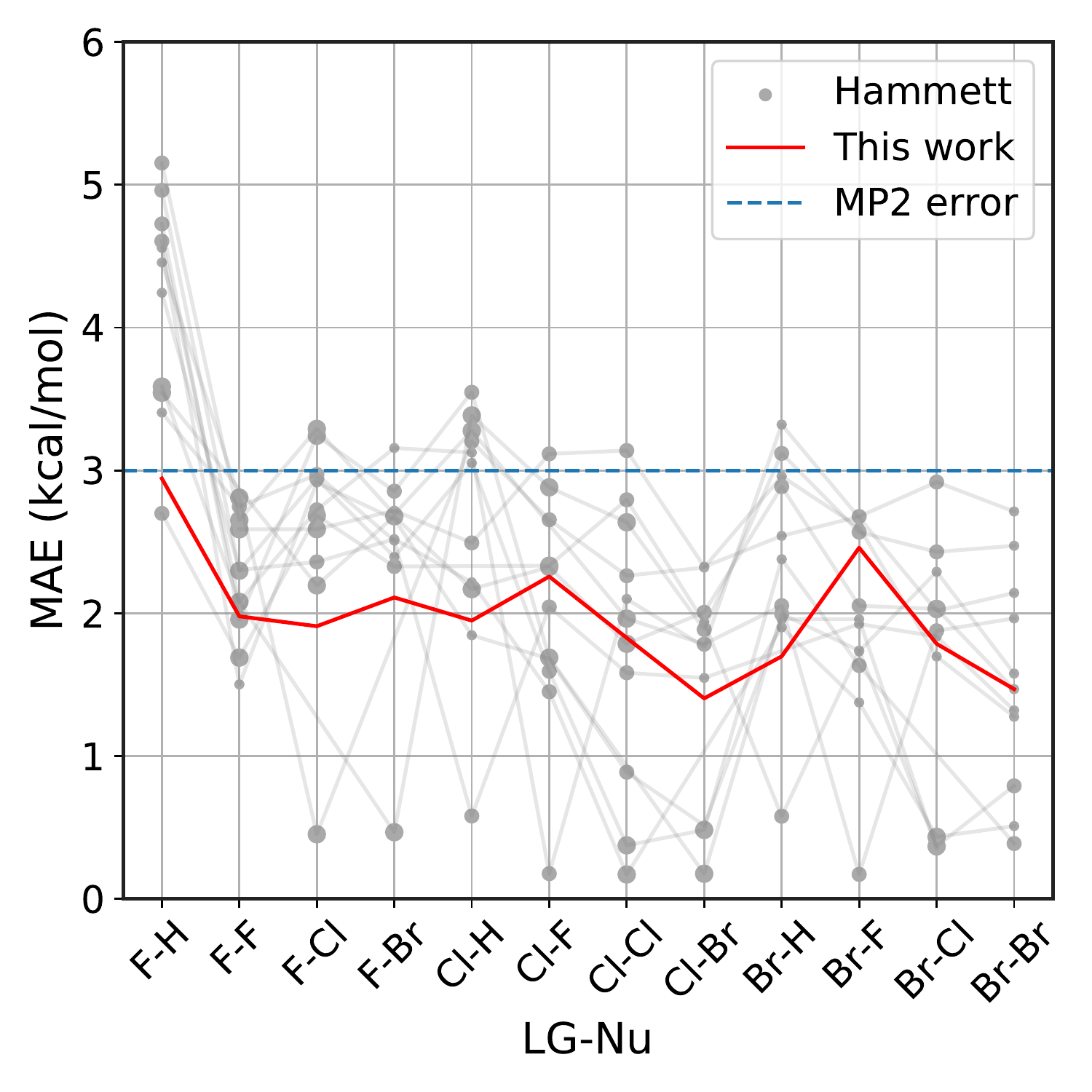}
    \caption{Accuracy of our model with the respect to the original Hammett approach. For each reaction, we show the mean absolute error (MAE) obtained with our model (red line) and with the original Hammett model (gray dots), where each dot represents a different choice for the reference reaction. The size of the dots is proportional to the size of the training set for that data point. The blue dashed line corresponds to the estimated MP2 error. \cite{MP2} \cite{MP2error}. }
    \label{fig:MAE}
\end{figure}

The improvement obtained with our method can be easily seen in figure~\ref{fig:MAE}. 
Here we present the Mean Absolute Error (MAE) for the prediction of the activation energy across all the reactions considered. 
The red line shows the MAE of our model,  while the gray dots the ones of the original model. 
For each reaction there are eleven dots, one for every different reference reaction.
The thin gray lines connect the results obtained with the same reference. 
The size of each dot is proportional to the number of common set of substituents between the reference reaction and the one being predicted. 
Finally, the dashed blue line shows the estimated error of the MP2 method \cite{MP2} \cite{MP2error}.

Our method outperforms the classic Hammett approach in the vast majority of the cases. 
For only a few reactions the original model can give better results but there is no single choice of reference that shows a consistently smaller MAE. 
Our method averages out the error obtained from the selection bias of the reference and gives a consistent prediction across all reactions, comparable in accuracy to the underlying MP2 method \cite{MP2error}.

The original method is highly susceptible to overfitting and numerical noise, as shown by the fact that small errors correspond mostly to medium size dots: few data points (small dots) lead to an unreliable fit, while too many (big dots) can make the model too rigid to be reliably transferable. 
This is especially evident for the two leftmost reactions (F-H and F-F), were the bigger data set are described very poorly by the original model. 
This can give MAE of up to 5.2\,kcal/mol, while our model has an error of 3.2\,kcal/mol at most.

As discussed in section~\ref{method:Hamm}, the original Hammett approach can get up to $N_R!$ different set of parameters, which for the 12 reactions considered here is in the order of $10^8$. 
The results shown in figure~\ref{fig:MAE} are obtained from a regression that considers only the reference reaction and the one for the prediction, so stopping the procedure after only two $\rho$ and a subset of $\sigma$ have been assigned. 
The factorial scaling of the extensive search makes it prohibitively expensive to find the best set of parameters.

\subsubsection{Decomposition of $\sigma$ for S$_N$2}\label{results:sigmadec}
The non-aromatic molecules we considered have four substituents attached to two different carbons atoms: two on the one involved in the reaction, from now on denoted as C1, and two on a carbon atom connected to C1 by a single bond, from now on denoted as C2. 
The molecular $\sigma$ for each set of substituents depends on all four groups and their position.
Via categorical regression, described in the Supplementary Information, it is possible to separate the individual contributions $\tilde{\sigma}$ and express the overall $\sigma$ as a linear combination. 

\begin{figure}[h!]
	    \centering
    \includegraphics[scale=0.32]{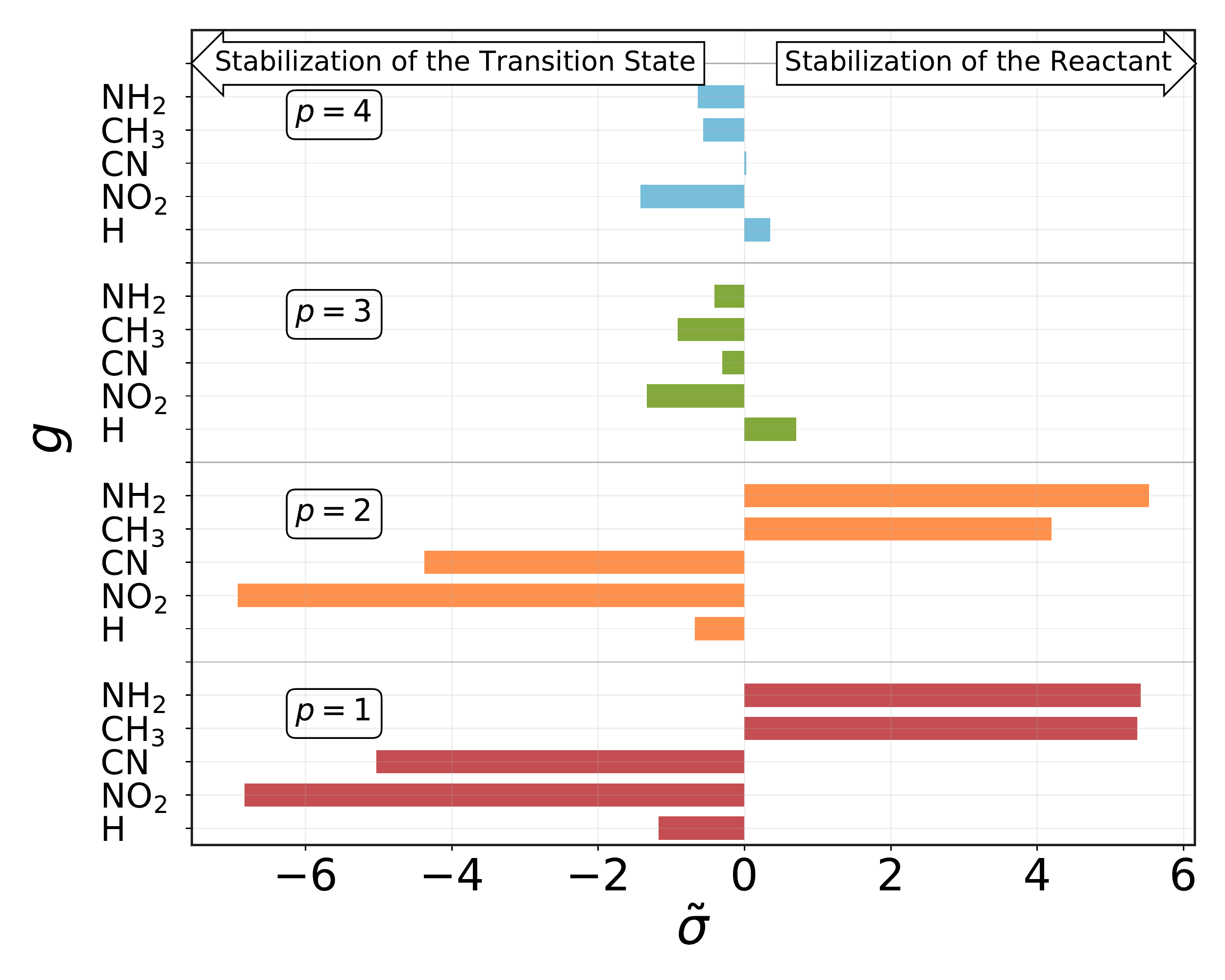}
    \caption{Contribution of each pair of group $g$ and position $p$ to the molecular $\sigma$, as obtained from the dummy encoding. Positive contributions give larger $\sigma$, resulting in higher activation energies, while negative contributions lead to a lowered barrier.}
    \label{fig:DummyParameters}
\end{figure}

The results of the decomposition are reported in figure~\ref{fig:DummyParameters}.
Each horizontal bar corresponds to one single-substituent $\tilde{\sigma}$ and the colors are used to distinguish the four positions: red and orange for positions 1 and 2, on C1, and green and blue for positions 3 and 4, on C2 (crf. figure~\ref{fig:compareEa}). 
The plot shows that the contributions given by positions 1 and 2 are almost identical. 
This makes sense chemically, since these two positions are nearly equivalent by symmetry (the molecule is chiral) and thus must have very similar effect on the reactivity of the molecule. 
The same is true for positions 3 and 4, although their absolute values of $\tilde{\sigma}$ are much smaller with respect to positions 1 and 2. 
Again, this follows chemical intuition, as these positions are further away from the reacting centre and their effect is dampened. 
These two  properties of $\tilde{\sigma}$ are not imposed at any point during the procedure, but they emerge by themselves.

The sign of the single substituent constants can be interpreted in the following way: if the reaction constant $\rho$ is positive, a substituent with a negative substituent constant $\sigma$ will give a lower activation energy than the reference substituent, and vice versa for positive $\sigma$. In our case, $\rho > 0$ for all reactions, so it possible to correlate the single substituent constants with the inductive effect. The electron withdrawing power of the groups considered goes as 
\begin{align*}
\text{-NO}_2 > \text{-CN} > \text{-H} > \text{-CH}_3 > \text{-NH}_2   
\end{align*}
Groups with negative values are electron withdrawing, while those with positive values are electron donating.
This again make sense chemically since the transition state of an S$_{\text{N}}$2 reaction is known to be negatively charged, and benefits more from a substituent that can remove electron density from the reacting centre. 
This chemical aspect, as well as the one regulating the magnitude of the substituents' effect depending on the position, is not imposed by the model but shows up naturally during the procedure.

\begin{figure}[h!]
	    \centering
    \includegraphics[scale=0.32]{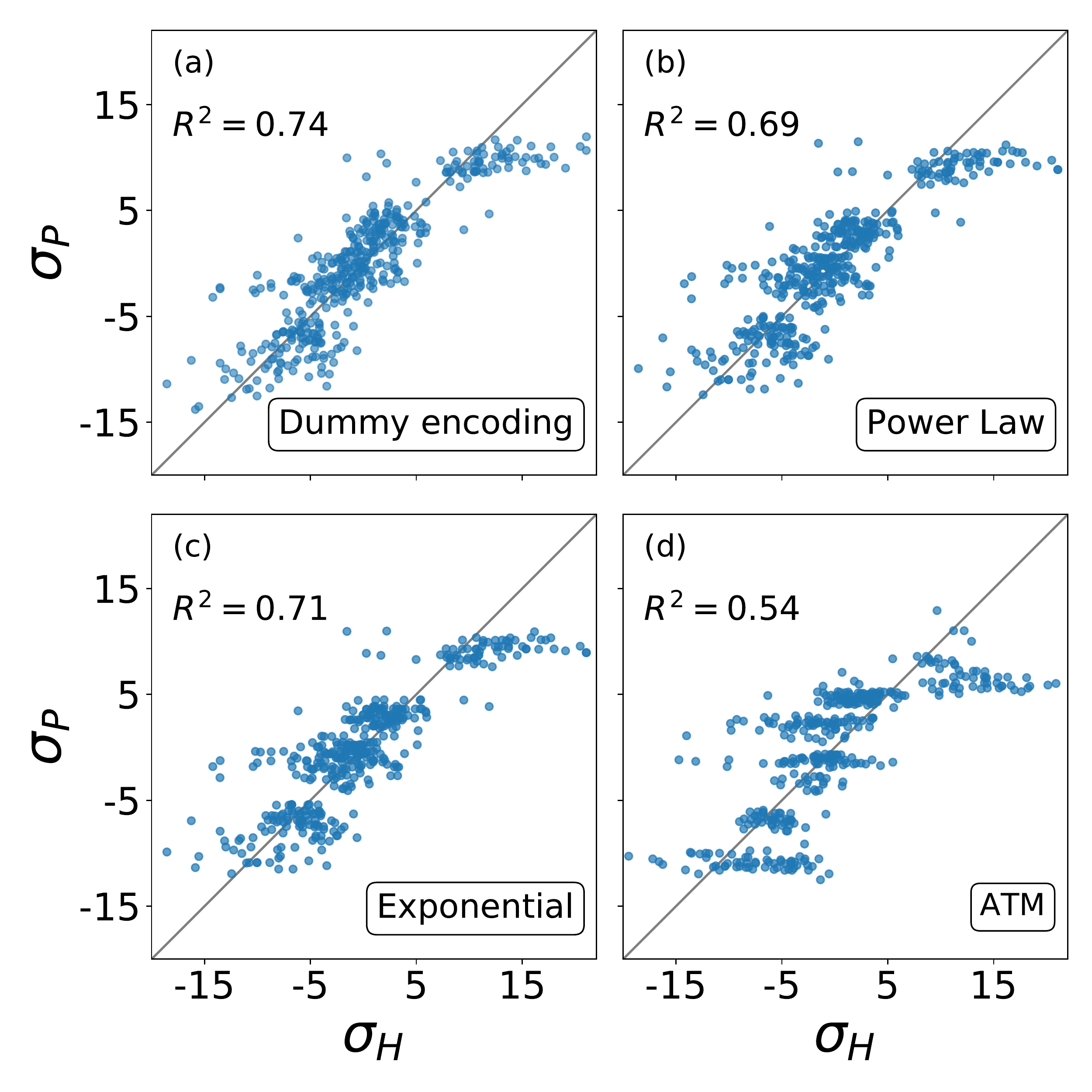}
    \caption{Correlation between the ${\sigma}$ obtained from the revisited Hammett and the ones obtained from: a) the dummy encoding, b) the power law function, c) the exponential function, d) the three body Axilrod-Teller-Muto function. Each panel also shows the $R^2$ of the correlation.}
    \label{fig:4panels}
\end{figure}

Although the single substituents constant obtained by the categorical regression depend on both their position and chemical composition at the same time, the results of this model indicate that these two can be further separated. 
We expressed the position dependence as the spatial separation from the reaction center, using a distance decaying function - we tested an exponential and power law one - that scales scales the electron withdrawing/donating effect of the substituent.
The latter is given by a constant which depends only the chemical composition.

The effects of interactions between different substituent on the molecular substituent constant can be modelled by a three-body term, as the Axlirod-Teller-Muto potential.

The results from these decompositions of the substituent constants are shown in figure~\ref{fig:4panels}. 
Here each scatter plot reports the correlation between the molecular ${\sigma}$ and the single-substituent ones, obtained with four different prediction methods: (a) categorical regression via dummy encoding, (b) power law function, (c) exponential function and (d) Axilrod-Teller-Muto (ATM) function.
Each panel shows the $R^2$ of the relative fit.

For each of these models, the number of parameters required depends on the number of substituent groups $N_G$ considered and the number of positions $N_G$ on the molecular backbone.
For our S$_N$2 datset, $N_G = 5$ (-H, -NO$_2$, -CN, -NH$_3$, -CH$_3$) and $N_P = 4$ (R1, R2, R3, R4) (crf. figure~\ref{fig:compareEa}).

The dummy encoding shown in plot (a) requires a total of $N_P N_G$ parameters, one for each group-position pair, so 20 for this data set. 
This approach has the great advantage of being independent from the backbone of the molecules, since it is sufficient to label each position and group. 
Including a new position or group in the data set would increase the number of parameters needed by $N_G$ (5) and $N_P$ (4) respectively. 
The prediction of the dummy encoding is good for most of the $\sigma$ in the set, showing some deviation only for values at the edge of the range.

For the exponential function and the power law in panels (b) and (c), the number of parameters required is $N_G + 1$, one for each group plus an additional one to regulate the distance decay. 
For our data set, this means six parameters. 
In this case, it is necessary to know the geometry of the molecular skeleton, which can be easily obtained. 
In terms of scalability, adding one more group increases the number of parameters by one, while for a new position it is only necessary to evaluate its distance from the reaction centre. 
The results obtained by these two functions are very similar to ones from the dummy encoding, but require significantly fewer parameters: in our case we go down from 20 to 6.

The Axilrod-Teller-Muto function shown in panel (d) takes into account the interaction between any two different groups in different positions on the molecule.
This requires a total of $N_G + (N_G^2 + N_G)/2 +1$ parameters: one for each group, one for every unique pair, and an additional one for the distance decay. 
For our data set, this brings us back to 20, as for the dummy encoding. 
For the ATM approach it is necessary to know the exact geometries of every molecule in order to calculate the distances and angles between different groups and positions. 
Extending the data set to a new group increases the parameters' cost by $1 + N_G$, i.e. 6. 
Including the interaction between groups and positions removes the simple additivity of single-substituents $\bm{\tilde{\sigma}}$ and actually worsens and the prediction.

Overall, fig.~\ref{fig:4panels} shows that the molecular substituent constants: (i) can be described quite well with only $N_G + 1$, i.e. 6 parameters, and (ii) show physical additivity.

\subsubsection{Comparison with Machine Learning for S$_N$2} \label{results:ML}
We compared the performance of our method with a Kernel Ridge Regression machine learning model.
We used a one-hot encoding representation, where each molecule is described with fingerprint-like string that depends on the functional groups present.
This representation was chosen because it contains no exact structural information, i.e. no cartesian coordinates, just like the categorical regression, making the comparison more fair as the two models work with the same information.
The machine was trained on both the activation energies and the residuals of the prediction from our revisited Hammett.
The latter approach is called Delta-Machine Learning, and uses as a baseline the predictions obtained from our $\alpha$-Hammett method, described in section~\ref{method:Hamm} and section~\ref{results:sigmadec}, where the substituent constants are obtained from a linear combination of single substituent contributions that are scaled by a distance decaying function. 
We choose this method as a baseline because it gives better predictions starting from smaller training set and because its residuals are more consistent, thus easier to learn.

\begin{figure}[h!]
	    \centering
    \includegraphics[scale=0.50]{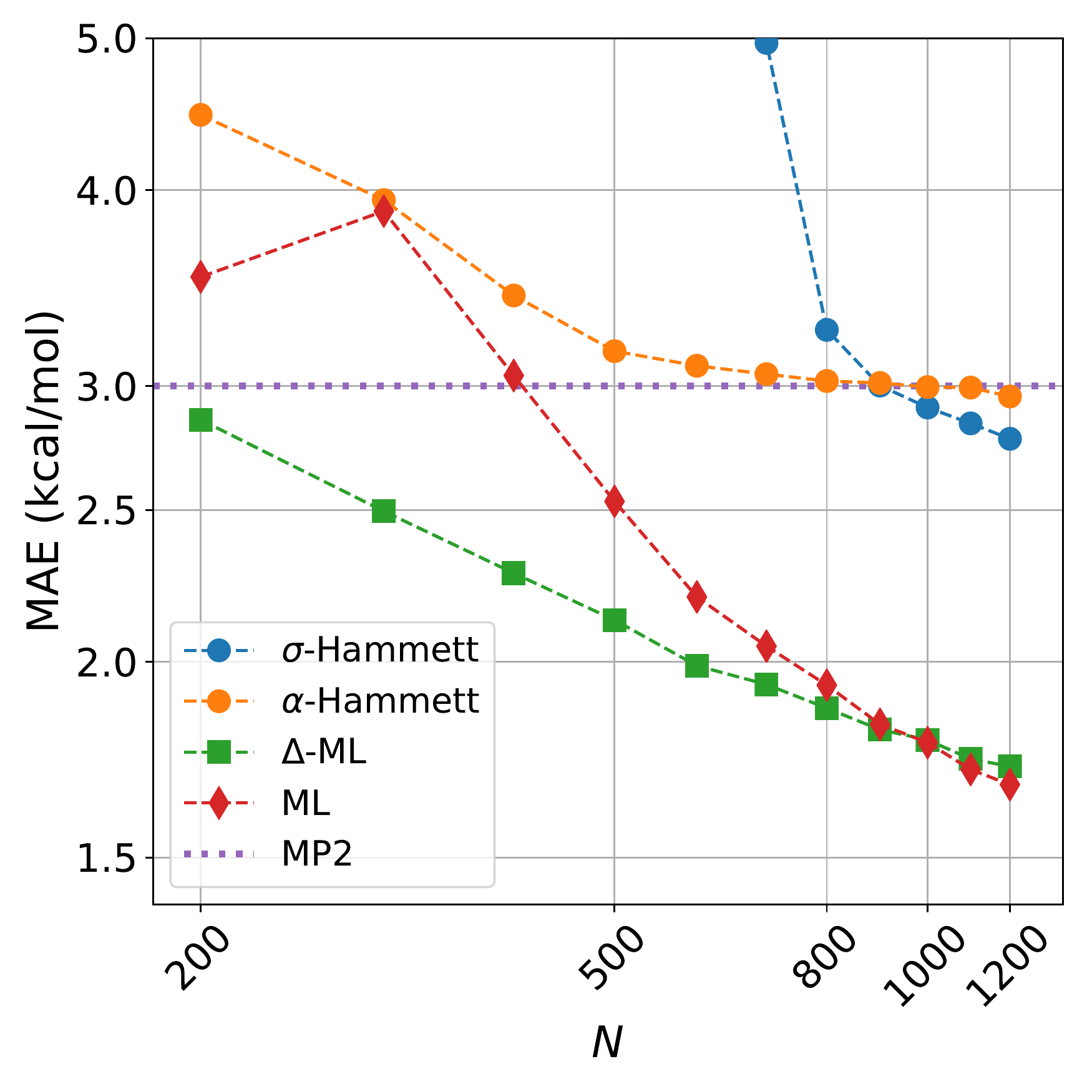}
    \caption{Learning curves for the activation energies with different methods. The circles are obtained by the Hammett model, where the $\sigma$ are calculated globally for the blue line ($\sigma$-Hammett) and additively for the green line ($\alpha$-Hammett). The diamonds and squares are given by Machine learning and $\Delta$-ML respectively, the baseline for the latter is $\alpha$-Hammett.}
    \label{fig:LC}
\end{figure}

The comparison of different methods is shown in figure~\ref{fig:LC}. Here we report different learning curves, which show how the performance of each method improves as the training set size increases.

 For a small training set, only some reactions and set of substituents can be sampled, giving values of $\rho$ that are highly influenced by random noise. 
 For the $\sigma$-Hammett model, this generates a set $\{\sigma\}$ that poorly reflects the true substituents' effect and gives very high prediction errors. 
 This method shows significant improvement with the increase of the training set size, and for 900 training points it recovers the accuracy of the reference method, MP2. 
 Using the complete data set recovers the accuracy shown in figure~\ref{fig:MAE}.
 
 The $\alpha$-Hammett method already gives errors below 5\,kcal/mol for only 200 training points and quickly converges to the accuracy of the underling level of theory. 
 The flattening out of the learning curve is due to the difficulty of this decomposition to describe accurately the values of $\sigma$ at the edges of the spectrum, as also shown in figure~\ref{fig:4panels}.
 
 The ML and $\Delta$-ML methods converge towards the same error, however the latter's learning curve has a significantly lower offset.
 This means that our method can also be used to speed up the learning of the target property at the cost of a very quick and inexpensive initial treatment of the data.
 The two learning curves converge at around 800 data points, where the baseline for the due $\Delta$-ML flattens out.
 Beyond this point, both methods just learn the MP2 error.

\section{Conclusion}
We developed a new method for calculating Hammett parameters $\rho$ and $\sigma$ that is generalized to include non-aromatic molecules and reactants with multiple substituents. 
We show that substituent effects are largely additive in this scenario as long as no resonance occurs.
In addition, for the $S_N2$ reaction space, Hammett $\sigma$ values can be explained by chemical composition and distance to the reaction center alone. 
This connects to the established view regarding the Hammett $\sigma$ values as a measure of the inductive effect and reduces the number of free parameters in our model. 

Moreover, we present a method to combine quantum chemical reference energies from several reactions into one reliable set of Hammett parameters. 
This allows to reduce the number of calculations required for real-world applications of Hammett's empirical relationship. 
Additionally, this reduces the risk of over-fitting towards one specific reaction which we demonstrate to be a significant problem with the original formulation. 

We tested this method on two different experimental data set and on a computational one and showed the improvement in both prediction quality and reliability. 
This method also provide an excellent baseline for $\Delta$-ML method, effectively allowing to cut down the size of training set.

With this improved method, Hammett's empirical formula can be employed as a guideline in reaction design without the need of extensive experimental or quantum chemical data set. 
We rather advocate for diverse data from many different reactions but common molecular skeletons, which then can be combined into one model following our approach. 
We demonstrated on our data set that this reaches the same accuracy of the underlying method of quantum chemical calculations. 
This way, Hammett's original idea can be used to uncover trends in reaction energies which are less affected by the systematic error of any quantum chemical method for different molecules, thus making a larger chemical space accessible.

\textbf{Acknowledgments.} \\
We acknowledge support by the European Research Council (ERC-CoG grant QML) as well as by the Swiss National Science foundation (No.~PP00P2\_138932, 407540\_167186 NFP 75 Big Data, 200021\_175747, NCCR MARVEL).
This work was supported by a grant from the Swiss National Supercomputing Centre (CSCS) under project ID s848.
Some calculations were performed at sciCORE (http://scicore.unibas.ch/) scientific computing core facility at University of Basel.

\bibliographystyle{achemso}
\bibliography{main.bib}

\end{document}


\section{Method}
\subsubsection{The original Hammett procedure}\label{method:originalHamm}
The Hammett equation was originally intended only for reactions occurring on simple aromatic molecules that have only one substituent on the ring. 
However, the equation itself contains no assumptions on the structure of the molecule.
Due to its linear nature, this equation can be applied to any data set and property $P$ where: (i) the ordering of the substituents with respect to $P$ is mostly stable across all reactions, (ii) the set of values for the property $P$ correlates linearly for any two reactions. 
The first condition is necessary to have one unique set of substituent constant for every reaction, the second allows to calculate $P$ using only a single multiplicative factor $\rho$.

\subsubsection{Hammett revisited}\label{method:newHamm}
The equilibrium constant can be expressed as a function of the free energy difference between product and reactant. The transition state theory extends this formulation to the kinetic constant by assuming a quasi-chemical equilibrium between transition state and reactant, thus using the free energy difference between these two. Both constants can be expressed as:

\begin{equation}
    K \propto \exp \left [{\frac{-\Delta G}{RT}}\right ]
    \label{deltaG}
\end{equation}

Thanks to eq~\ref{deltaG}, we can replace the $\log K$ in the Hammett equation  with a free energy difference $\Delta G$ or a potential energy difference, since it meets the conditions imposed by the Hammett equation presented above.
The logarithm of the kinetic constant can be replace by the activation energy $E_a$, giving:
\begin{equation}
    E_a(s,r) - E_0(r) \simeq \rho(r) \sigma(s) 
    \label{NewHamm}
\end{equation}
where $r$ is one of the $N_R$ reactions, $s$ one the $N_S$ set of substituents and $E_0$ is the activation energy for the unsubstituted molecule.

We first evaluate the set of reaction constants $\{\rho\}$. If we compare the activation energies of any two different reactions $r_i$ and $r_j$ which share common set of substituents, we obtain the following system.
\begin{equation}
    \begin{split}
    E_a(s,r_i) - E_0(r_i)  &\simeq \rho(r_i) \sigma(s) \\
    E_a(s,r_j) - E_0(r_j) &\simeq \rho(r_j) \sigma(s)
    \end{split}
\end{equation}

Dividing the first equation by the second one gives:
\begin{equation}
     E_a(s,r_i)  \simeq \frac{\rho(r_i)}{\rho(r_j)} \left [ E_a(s,r_j) - E_0(r_j) \right ]  + E_0(r_i)  
     \label{DivideHamm}
\end{equation}

 Linear regression of energies allows to express the ratio of the two $\rho(r_i)$ and $\rho(r_j)$ as the slope $m$ of the line, yielding:
 \begin{equation}
     m \rho(r_j) - \rho(r_i) = 0
 \end{equation}
 This gives a system of $N_R^2 - N_R$ equations that can be solved to obtain the $N_R$ values of $\rho$.
 We made use of a robust regressor (Theil,~H. A rank-invariant method of linear and polynomial regression analysis.   {I}. \emph{Nederl. Akad. Wetensch., Proc.} \textbf{1950}, \emph{53}, 386--392.) to minimize the impact of strong outliers on the final values.
 The initial value of one of the reaction constants $\rho$ was fixed to 1 to avoid trivial. This is the only source of bias in the procedure, its effects are discussed below.
 We treated each $E_0(r)$ as a model parameter and set it to the median of all the activation energies available for the reaction $r$. 
 This is done in order to reduce the dependence of the model on only $N_R$ calculations.
 
 Once the $\{\rho\}$ is defined, the substituent constants are calculated as:
 \begin{equation}
     \sigma(s) \coloneqq \frac{1}{R} \sum_{r=1}^{N_R} \frac{E_a(s,r) - E_0(r)}{\rho(r)}
     \label{eq:calcsigm-global}
 \end{equation}
 
It is possible to further improve the parameters by noting that for a fixed reaction $\widetilde{r}$, eq~\ref{NewHamm} can be rewritten as:
\begin{equation}
     E_a(s,\widetilde{r}) = \widetilde{m} \left [ \rho(\widetilde{r})  \sigma(s) \right ] + E_0(\widetilde{r}) + \widetilde{q}
 \end{equation}
 
 Via least squares regression it is possible to find values of $\widetilde{m}$ and $\widetilde{q}$, which, for a perfect correlation, should be equal to $1$ and $0$, respectively. 
 $\rho$ and $E_0$ can be tuned to improve the correlations according to:
 
\begin{equation}
    \begin{split}
        &\rho(\widetilde{r}) \coloneqq \rho(\widetilde{r}) - 1 + \widetilde{m} \\ 
        &E_0(\widetilde{r}) \coloneqq E_0(\widetilde{r}) + \widetilde{q}
    \end{split}
    \label{fixrhoe0}
\end{equation}

\subsubsection{Decomposition of $\sigma$} \label{method:sigmadec}
The substituent constants obtained from eq~\ref{eq:calcsigm-global} are molecular properties, which describe the effect of the entire set $s$ of substituents. 
By denoting each substituted position on the molecule by the index $p$ and each substituent group (e.g. NO$_2$) by the index $g$, we highlight the dependency of each $\sigma$ as $\sigma(s) = \sigma(\{ g_p\})$, where by $g_p$ we indicate the group $g$ to be in position $p$. 
If $N_P$ is the total number of positions $p$ on the molecule, and $N_G$ the total number of substituent groups $g$, the maximum number of set $s$ is $N_G^{N_P}$. 
However, each molecular $\sigma$ depends only on $N_P$ terms at most.
The overall $\sigma(s)$ can be expressed as a linear combination of these $N_P$ terms:

\begin{equation}
\sigma(s) = \sum_{p=1}^{N_P} \tilde{\sigma}(g_p)
\label{LCsigm}
\end{equation}
The $\tilde{\sigma}(g_p)$ are the single-substituent sigmas. 
They are independent from one another and can be determined via categorical regression using a dummy encoding.
In this, fingerprint-like representation, each molecule in the data set is described by a vector of $N_P N_G$ values, representing all the possible combinations of position and group. 
All the elements are zeros, except for the ones corresponding to the group-position pairs present in the molecule. 
These vectors are then stacked into a matrix $\mathbb{A}$ which is then used to solve the linear system

\begin{equation}
	\mathbb{A} \bm{\tilde{\sigma}} = \bm{\sigma}
	\label{dummyeq}
\end{equation}

This type of decomposition reduces the number of parameters needed to describe the substituents from $N_G^{N_P}$ to $N_G N_P$ and allows to predict values of $\sigma(s)$ for set of substituents for which no data is available.
However, these $\tilde{\sigma}(g_p)$  still depend on both the position and the group, meaning that the same group will have a different value depending on its position on the molecules. 
While this is chemically sound, it limits the transferability of the model.
To separate the effect of the group $g$ from the one of the position $p$, we replace the dependence on the latter by distance decaying function that scales the single-substituent effect.
This is way the energy difference is modelled after the electronic density.
Here an exponential decaying push/pull effect is given by electron withdrawing and electron donating group, respectively.

This can be modelled by the following functions:

\begin{align}
    \sigma(s) &= \sigma(\{ g_p\}) = \sum_{p=1}^{N_P} \alpha(g) \exp{\frac{-d_{p}}{\tau}} \label{eq:expfunc}\\ 
    \sigma(s) &= \sigma(\{ g_p\}) = \sum_{p=1}^{N_P} \frac{\alpha(g)}{d_{p}^\tau} \label{eq:powerlaw}
\end{align}

where $\alpha(g)$ is a parameter which depends only on the group $g$, regardless of its position on the molecule, $d_{p}$ is the distance between the position $p$ and the reacting centre on the molecule, and $\tau$ is a parameter of the model which regulates the distance decay of the inductive effect. 
$\alpha(g)$ is determined by a linear regression while the optimal $\tau$ can be found by a scan. 
This approach further cuts down the number of parameters required by the model to describe the substituents from $N_PN_G$ to $N_G + 1$. 
It requires geometrical information on the backbone of the molecule, which is easily obtainable.

Eq~\ref{LCsigm},~\ref{eq:expfunc} and~\ref{eq:powerlaw} all neglect the interactions between different group-position pairs. These could be modelled by three body terms such as the Axilrod-Teller-Muto (Axilrod,~B.~M.; Teller,~E. Interaction of the van der Waals Type Between Three   Atoms. \emph{The Journal of Chemical Physics} \textbf{1943}, \emph{11},   299--300.) potential form:
\begin{equation}
	V_{ijk} = \frac{1 + 3 \cos \gamma_i\cos \gamma_j\cos \gamma_k }{r_{ij}r_{jk}r_{ik}}
	\label{eq:ATM}
\end{equation}
 In this case, V is is not a potential, but it keeps the same functional form and includes distances and angles between any two group-position and the reacting centre. 
 This can be used to describe the residuals of the previous fit by including many-body effects. 
 The added flexibility comes at the cost of $(g^2 + g)/2$ additional parameters, one for each possible substituent pair.

\subsubsection{Dependence on the reference reaction}
\begin{figure}[h!]
	    \centering
    \includegraphics[scale=0.4]{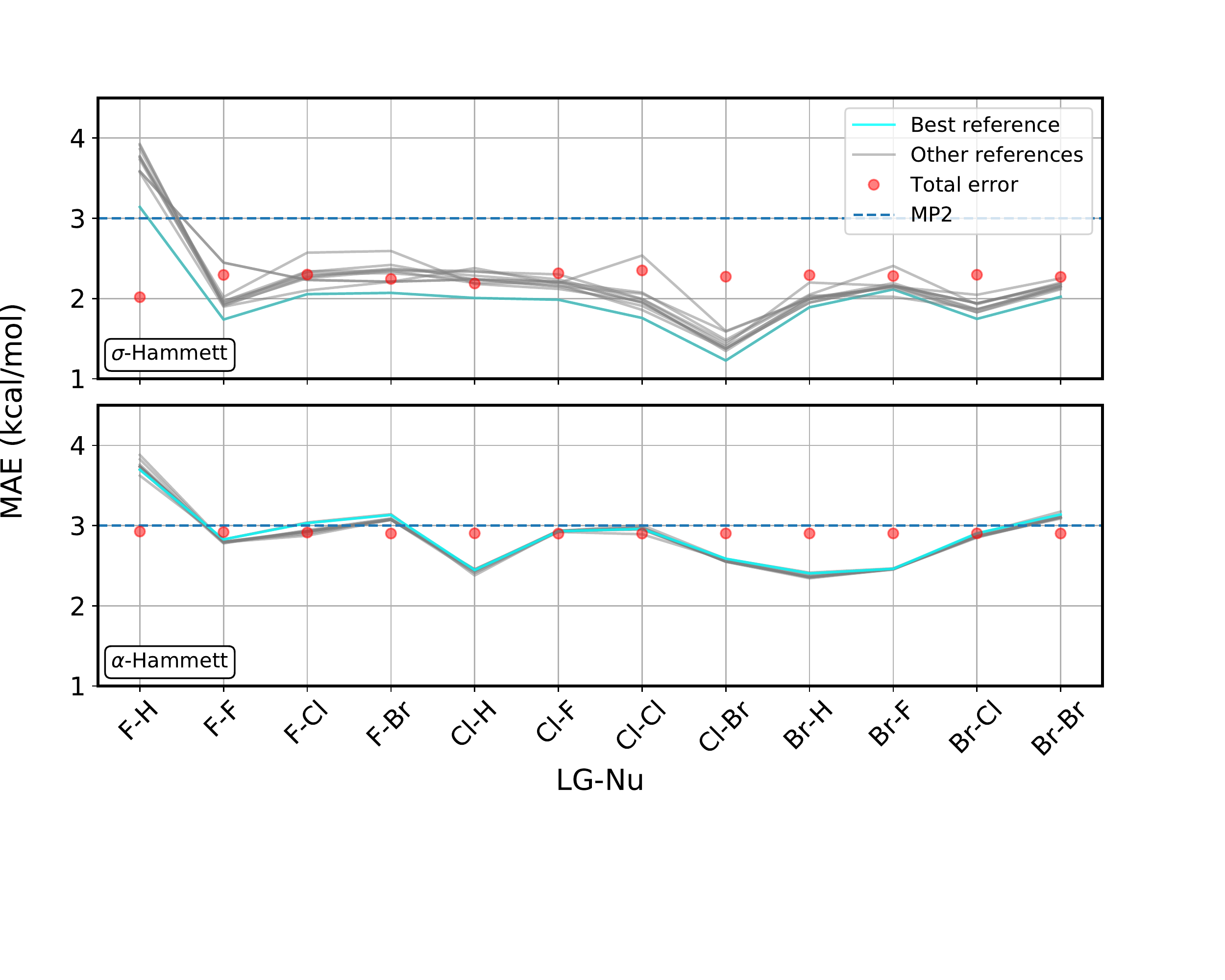}
    \caption{Influence of the reference reaction on the Mean Absolute Error (MAE) of the prediction of activation energies. Red circles report the overall MAE when the reaction listed on the x-axis is used as a reference. The gray lines, one for each different reference, show the error on the prediction on each specific reaction}
    \label{fig:SI}
\end{figure}

As discussed above, it necessary to initially set on of the reaction constants $\rho$ to 1, in order to avoid trivial solutions.
This is the only source of bias in our model and its effect is observed to be limited.
For the experimental data sets described in section III.1, the effect of the reference's choice is shown in figure 1.
For the computational S$_N$2 data, we show the influence of the reference choice in figure~\ref{fig:SI}.

The two panels show the Mean Absolute Error (MAE) of the prediction of activation energies.
For the top panel, the substituent constants are obtained from eq~\ref{eq:calcsigm-global}; we named this method $\sigma$-Hammett.
In the bottom panel, the substituent constants are obtained from the sum of individual contributions with a power-law distance decay, as calculated from eq~\ref{eq:powerlaw}; we named this method $\alpha$-Hammett.

Each gray line corresponds to a different choice for the reference reaction, out of the 12 listed on the x-axis, and shows how the MAE changes across the reaction space.
In each panel, we highlighted in blue the one that gives the best overall prediction.
The red circles show the total error, i.e. across all the 12 reactions, for each reference indicated indicated on the x-axis.
These results are compared to the accuracy of the MP2 method, shown by the dashed line.

These plots shows how the overall prediction, given by the red circles is only partially affected by the reference bias, especially for the $\alpha$-Hammett model.
Additionally, the gray lines are all very close to each other, meaning that even the description on smaller subset of the data remains mostly consistent regardless of the reference reaction chosen.

The $\alpha$-Hammett model gives a worse prediction, by about 0.75 kcal/mol on average, but it almost completely negates the effect of the reference bias.

\subsubsection{Machine Learning} \label{method:ML}
The activation energies can also be obtained from Machine Learning. In this work we use Kernel-Ridge Regression, for which the property of interest $y$ of a molecule $\widetilde{X}$ can be predicted as:
\begin{equation}
	y (\widetilde{\mathbf{X}}) \simeq \sum_{i}^{N} \alpha_i k(\widetilde{\mathbf{X}}, \mathbf{X}_i)
	\label{KRR}
\end{equation}
where $i$ runs over all the molecules in the training set, $\alpha_i$ are regression coefficients and $k(\mathbf{X}, \mathbf{X_i})$ is a kernel function. In this work, we used a Laplacian kernel, where each element $j,i$ is given by:
\begin{equation}
k_{j,i} = \exp\left( - \frac{\| \mathbf{A}_j \mathbf{B}_i \|_1}{w}  \right)
\label{LaplKern}
\end{equation}
where $\mathbf{A}_j$ and $\mathbf{B}_j$ are representation vectors and $w$ is the kernel width. The regression coefficients $\alpha_i$ can be calculated as:
\begin{equation}
	\alpha_i = ( \mathbb{K} + \lambda \mathbb{I}) ^{-1} y
\end{equation}
where $\lambda > 0$ is a hyperparameter used as a regularizer and $\mathbb{K}$ and $\mathbb{I}$ are the kernel matrix and identity matrix respectively. As representation $\mathbf{X}$ we used the one-hot encoding described in sec.~\ref{method:sigmadec}. 
In this case, the string describes not only the set of substituents, but also the reaction being considered, and for this reason it contains R extra characters, one for each reaction in the data set. 
This type of machine learning algorithm is used to either predict directly the activation energies or to learn the residuals of the Hammett regression using Delta Machine Learning (Ramakrishnan,~R.; Dral,~P.~O.; Rupp,~M.; von Lilienfeld,~O.~A. Big Data Meets   Quantum Chemistry Approximations: The $\Delta$-Machine Learning Approach.   \emph{Journal of Chemical Theory and Computation} \textbf{2015}, \emph{11},   2087--2096). 
The latter works on the assumption that learning the target property from a smoother surface is easier, and thus requires fewer training points to reach high accuracy. 

\bibliography{main}